# Band-to-band tunneling in a carbon nanotube metal-oxide-semiconductor field-effect transistor is dominated by phonon assisted tunneling

*Siyuranga O. Koswatta,[†,*] Mark S. Lundstrom,[†] and Dmitri E. Nikonov[‡]*

† School of Electrical and Computer Engineering, Purdue University, West Lafayette, Indiana 47906

‡ Technology and Manufacturing Group, Intel Corp., SC1-05, Santa Clara, California 95052

* Corresponding author. E-mail: koswatta@purdue.edu

**Abstract** – Band-to-band tunneling (BTBT) devices have recently gained a lot of interest due to their potential for reducing power dissipation in integrated circuits. We have performed extensive simulations for the BTBT operation of carbon nanotube metal-oxide-semiconductor field-effect transistors (CNT-MOSFETs) using the non-equilibrium Green's functions formalism for both ballistic and dissipative quantum transport. In comparison with recently reported experimental data (Y. Lu et al, *J. Am. Chem. Soc.*, v.

128, p. 3518-3519, 2006), we have obtained strong evidence that BTBT in CNT-MOSFETs is dominated by optical phonon assisted inelastic transport, which can have important implications on the transistor characteristics. It is shown that under large biasing conditions two-phonon scattering may also become important.

Device power dissipation has become a major challenge for the continued scaling of integrated circuits.[1,2] One important factor contributing to the overall power dissipation is poor sub-threshold (off-state) properties of modern transistors. This has led to large sub-threshold leakage currents, and it has limited the ability to scale the power supply voltage ($V_{DD}$), which has been the preferred way to decrease power dissipation in modern integrated circuits. The sub-threshold swing (*S*) for conventional transistors, which determines how effectively the transistor can be turned off by changing the gate voltage ($I_{DS}$ vs. $V_{GS}$), has a fundamental limit of $S = 2.3 \times k_B T / q$ mV/decade ≈ 60mV/dec at room temperature, where $k_B$ is the Boltzmann constant, *T* the temperature, and *q* the electron charge.[3] The need to maintain a certain ratio of on-current to off-current limits scaling of the power supply voltage, aggravating heat dissipation problems in modern high-performance circuits.

The aforementioned limitations on the off-state performance of conventional transistors equally apply to those based on carbon nanotubes (CNTs): both Schottky-barrier transistors (SB-CNTFETs) and metal-oxide-semiconductor field-effect transistors (CNT-MOSFETs) with doped source/drain contacts.[4,5,6,7] It has been experimentally observed that carrier transport in CNTs can be nearly ballistic.[8,9,10] High-performance CNT transistors operating close to the ballistic limit have also been demonstrated.[11,12,13] Recently, p-type CNT-MOSFETs with near ideal gate control and sub-threshold operation close to the fundamental limit of $S \approx 60$mV/dec at negative gate voltages ($V_{GS}$) have been reported.[14,15] At positive voltages, these CNT-MOSFETs begin to turn on again with a swing of 40~50mV/dec which is *smaller* than the conventional limit of 60mV/dec. This regime of operation is attributed to band-to-band tunneling (BTBT),[14,16] and should be distinguished from the ambipolar operation in SB-CNTFETs where the ambipolar branch is due to opposite type of carrier injection (electrons in the case of p-type FETs) from the drain Schottky contact.[6,17] Specialized device geometries including that based on BTBT have been actively investigated to obtain improved sub-threshold performance in CNT-MOSFET transistors.[18,19] In this paper, we present detailed theoretical simulations for the BTBT operation of CNT-MOSFETs using the non-equilibrium Green's functions (NEGF) formalism[20] for both ballistic and dissipative quantum transport. Comparing these results with the recently reported data in Ref. 15, we provide compelling evidence that the BTBT operation of CNT-MOSFETs is strongly affected by phonon-assisted tunneling.

The idealized device structure used in this study, shown in Figure 1, is a p-type CNT-MOSFET with wrap-around high-k (HfO$_2$ ~ $\kappa$ = 25) gate dielectric with thickness $t_{OX}$ = 2nm, doped source/drain regions ($N_{S/D}$ = 0.6/nm) with $L_{SD}$ = 40nm, and an intrinsic channel length of $L_{ch}$ = 30nm. A zigzag (16,0) CNT with diameter ~ 1.2nm and bandgap ~ 0.7eV is considered. The CNT-metal contacts at the ends of source/drain regions are assumed to be Ohmic, with perfect transparency for carrier transport. In experimental devices the Ohmic contacts with electrostatically induced doping is achieved through a back-gated geometry.[14,15,21] Although the device structure is idealized, the device parameters are in agreement with those in Ref. 15. Our simulations are performed for p-type CNT-MOSFETs in correspondence with experiments. Due to the symmetry of conduction and valence bands in CNTs,[22] the simulated *I-V* curves for n-type and p-type transistors are expected to coincide after reversing the polarity of voltages; and indeed we have verified this equivalence.

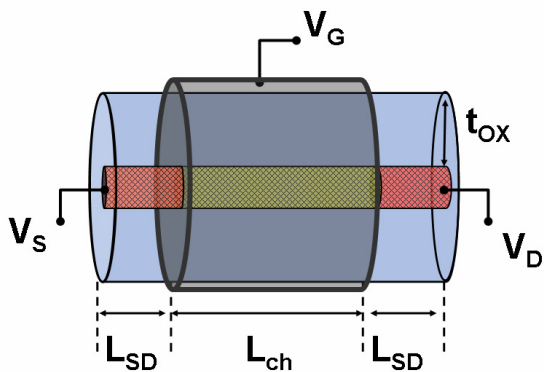

Figure 1. Simulated p-type CNT-MOSFET with doped source/drain regions and wrap-around gate electrode. See text for device parameters.

The NEGF calculations for dissipative transport are performed self-consistently with electrostatic simulations.[23] We use the nearest neighbor $p_z$ tight-binding Hamiltonian ($H_{p_z-TB}$) to describe the CNT electronic structure.[20,22] The simulations are carried out in the mode-space, considering transport through the first conduction and valence subbands ($E_{C1}$ and $E_{V1}$), respectively.[23,24] The difference of energy to the next highest subbands, i.e. $|E_{C2}-E_{C1}| = |E_{V2}-E_{V1}|$, is about 370meV so the transport through higher subbands can be neglected for typical biasing conditions. The retarded Green's function for the device under electron-phonon (e-ph) coupling is given by,

$$G^r(E) = \left[ EI - H_{p_z-TB} - \Sigma_S - \Sigma_D - \Sigma_{scat} \right]^{-1} \quad (1)$$

where $\Sigma_{S/D}$ are the self-energies for source/drain reservoirs and $\Sigma_{scat}$ is the self-energy for e-ph interaction determined using the self-consistent Born approximation.[25,26] All self-energy functions are energy dependent, and the energy indices are suppressed for clarity. The scattering self-energy is related to the in/out-scattering functions ($\Sigma_{scat}^{in/out}$) by,

$$\Sigma_{scat} = -\frac{i}{2}\left[\Sigma_{scat}^{in} + \Sigma_{scat}^{out}\right] \equiv -\frac{i}{2}\Gamma_{scat} \qquad (2)$$

where we have neglected the real part for simplicity. The electron/hole correlation functions, $G^{n/p}(E)$, are given by,

$$G^{n/p} = G^r \Sigma^{in/out} G^{r\dagger}. \qquad (3)$$

The in/out scattering functions $\Sigma^{in/out}$ in Eq. (3) have the contributions from the contact as well as e-ph interaction functions.[25,26] The diagonal elements of the correlation functions, $G^{n/p}(z,z,E)$, relate to the induced charge density on the CNT surface, $Q_{ind}(z)$, that is used in determining the self-consistent potential, $U$, from Poisson's equation in cylindrical coordinates,

$$\nabla^2 U(r,z) = -\frac{Q_{ind}(z) + N_D^+ - N_A^-}{\varepsilon} \qquad (4)$$

where, $N_D^+$ and $N_A^-$ are the ionized donor and acceptor doping densities, respectively. It should be noted that we use the Neumann boundary condition, $\vec{\nabla}U \equiv 0$, at source/drain ends, indicating charge neutrality in the doped reservoirs. The detailed treatment of the self-consistent procedure can be found in Ref. 23.

We include one-phonon and two-phonon scattering processes in our simulation. In/out scattering functions in Eq. (2) for one-phonon scattering with optical phonons of energy $\hbar\omega$ are given by,

$$\Sigma_{scat}^{in}(z,z,E) = R_{OP}(N_\omega+1)G^n(z,z,E+\hbar\omega) + R_{OP}N_\omega G^n(z,z,E-\hbar\omega) \quad (5)$$

$$\Sigma_{scat}^{out}(z,z,E) = R_{OP}(N_\omega+1)G^p(z,z,E-\hbar\omega) + R_{OP}N_\omega G^p(z,z,E+\hbar\omega) \quad (6)$$

where $N_\omega$ is the equilibrium Bose-Einstein distribution. The e-ph coupling parameter, $R_{OP}$, for optical phonon scattering in an $(n,0)$ zigzag CNT is, $R_{OP} = \hbar J_1^2 |M_{OP}|^2 / (2nm_C\omega_{OP})$, with $J_1$ = 6eV/Å, $|M_{OP}|$ calculated according to Ref. 27, and $m_C$ the mass of carbon atom. For the (16,0) CNT and the longitudinal optical (LO) phonon mode considered in this study, $\hbar\omega_{OP}$ = 195meV and $R_{OP}$ = 0.01eV$^2$, respectively. In Eqs. (5) and (6) the scattering functions are taken to be diagonal, which is the case for local interactions.[23,25,26] Two-phonon scattering is included similarly, with the phonon energy replaced by $\hbar\omega_{2\text{-}ph}$ = 360meV, corresponding to 180meV zone-boundary optical phonons, and the coupling parameter related to that of the one-phonon transition via,[28]

$$R_{e-2ph}(E) = R_{OP}\frac{2\Gamma_{OP}}{\Gamma_{tot}(E\pm\hbar\omega_{2ph}/2)}, \quad (7)$$

where $\Gamma_{OP}$ is the one-phonon broadening and $\Gamma_{tot}$ is the total broadening due to scattering and the source/drain reservoirs. Finally, the current through the device from node $z$ to $(z+1)$ in nearest neighbor tight-binding model is determined by,[25,26]

$$I_{z\to z+1} = \frac{4qi}{\hbar}\int\frac{dE}{2\pi}\left[H_{p_z-TB}(z,z+1)G^n(z+1,z,E) - H_{p_z-TB}(z+1,z)G^n(z,z+1,E)\right]. \quad (8)$$

We have performed self-consistent NEGF calculations for both ballistic ($\Sigma_{scat} = 0$) and dissipative ($\Sigma_{scat} \neq 0$) transport using the efficient numerical algorithms reported in Ref. 29. The simulation results are compared against the data reported in Ref. 15; the relevant experimental $I_{DS}$-$V_{GS}$ plot is reproduced in Figure 2 for the sake of completeness. In examining Figure 2 we can observe a few important features of the p-type CNT-MOSFET operation: 1) near ideal ($S \approx 60$mV/dec) sub-threshold behavior under conventional MOSFET operation for negative $V_{GS}$ biases, 2) larger on-currents for the conventional operation compared to BTBT regime, 3) $S \approx 50$mV/dec (< 60mV/dec conventional limit) is observed for BTBT transport at $V_{DS}$ = -0.01V. *The BTBT sub-threshold swing, however, degrades with increasing $V_{DS}$ biases.* 4) *The onset of BTBT occurs at a smaller $V_{GS}$ with increasing $V_{DS}$*, i.e., the onset $V_{GS}$ bias point moves left on Figure 2 with increasing $V_{DS}$. These experimental device characteristics will be compared against our computational results in order to elucidate the transport mechanisms.

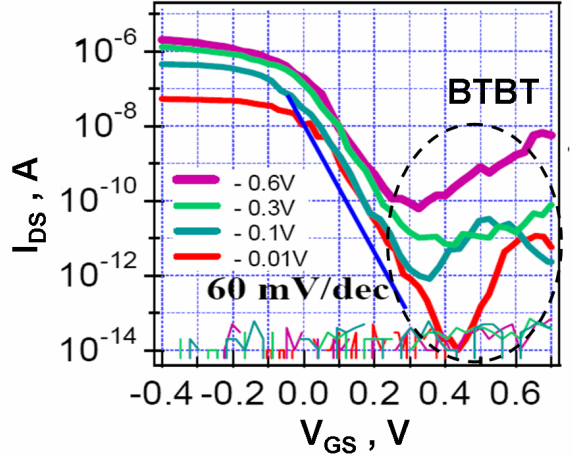

Figure 2. Experimental $I_{DS}$-$V_{GS}$ data at different $V_{DS}$ biases for the p-type CNT-MOSFET; reproduced from Ref. 15.

Ref. 16 presents a detailed study of the BTBT operation in a CNT-MOSFET, and here we summarize the transport mechanism responsible for it. Figures 3 (a) and (b) depict the two main mechanisms for BTBT transport in p-type CNT-MOSFETs with sufficiently short channel lengths. For such devices, due to longitudinal confinement inside the channel region, we observe quantized states in the conduction band. As shown in Figure 3(a), at sufficiently large positive gate biases these quantized states align with filled states of the valence band in the drain region, and direct (resonant) tunneling to the source region becomes possible. This alignment has a sharp onset, thus leading to steep sub-threshold slopes in $I_{DS}$-$V_{GS}$ characteristics.[16] This can be further understood by observing the hole Fermi distribution in the source region, noting that carrier conduction is equally explained through hole transport from source to drain. Here, bottom of the hole Fermi distribution in the source is cut off by the valence band edge, and the top is cut off by the conduction band edge in the channel. The result is a cooler current-carrying distribution, which leads to a smaller subthreshold swing. For moderate gate biases, as shown in Figure 3(b), direct tunneling through the conduction band states is prohibited, and the current will be dominated by thermionic emission of holes under the channel barrier in the valence band.

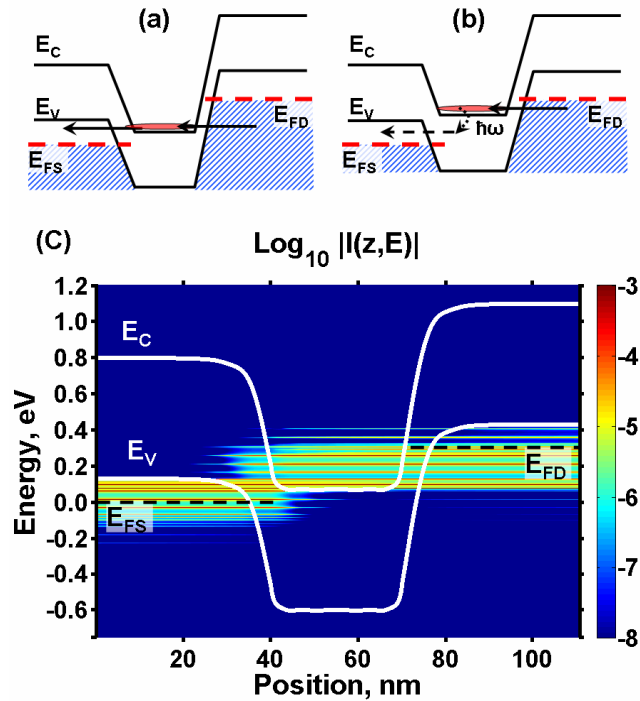

Figure 3. BTBT operation in a p-type CNT-MOSFET: (a) For large positive $V_{GS}$, direct tunneling of electrons from drain to source through the quantized conduction band states in the channel region, (b) For moderately positive $V_{GS}$ direct tunneling is prohibited, but inelastic tunneling is possible in the presence of optical phonons (dashed arrows), (c) NEGF simulation results for energy-position resolved current density spectrum (logarithmic scale) under 195meV LO phonon scattering at $V_{GS}= 0.4$V, $V_{DS} = -0.3$V.

If we take into consideration high-energy optical phonon scattering, another transport channel is possible via phonon-induced virtual states as shown by the dashed arrows in Figure 3(b). In this case, for BTBT to occur, only the phonon virtual state needs to align with the empty valence band states in the source, and the drain bias should be large enough to allow the quantized conduction band states to be filled by the drain. The existence of the two transport channels is inferred from energy-position resolved current density spectrum in Figure 3(c). Thus, it can be expected that in the presence of optical phonons the onset of BTBT transport will begin at a smaller gate bias, and the device *I-V* characteristics will be different from those expected in the ballistic case.[16]

Figure 4(a) shows the simulated $I_{DS}$-$V_{GS}$ characteristics under ballistic transport for the p-type CNT-MOSFET given in Figure 1. It is seen that the sub-threshold swing for regular MOSFET operation (negative $V_{GS}$ for p-type devices) is indeed at the theoretical limit of 60mV/dec, which is not surprising given the wrap-around gate geometry we have employed in our simulations. The on-current for regular operation is also larger than that for the BTBT operation. The overall device currents observed in our simulator, however, are larger than the experimental ones, which can be attributed to contact resistance in the experiment. The simulated sub-threshold swing in the BTBT regime is $S \approx 20$mV/dec at $V_{DS}$ = -0.01V. Under ballistic transport, the steep slope for BTBT regime does not degrade for large $V_{DS}$, contrary to the experimental results in Figure 2. More importantly, *the onset of BTBT transport moves to more positive gate voltages with increasing $V_{DS}$*, i.e. to the right, as shown in Figure 4(a). This behavior under ballistic transport is well understood and is attributed to the "charge pile-up", or, "floating body" effect.[6,30] As shown in Figure 3(b), under ballistic conditions at large $V_{DS}$ and moderate $V_{GS}$, the conduction band states are not aligned with the empty source states but are aligned with filled states in the drain. Carriers can tunnel into these states, resulting in a pile-up of electrons. Their potential causes the bands in the channel to move up, thus requiring an even larger positive gate bias before the onset of BTBT. As a result, the onset point in Figure 4(a) moves right to larger gate voltages with increasing $V_{DS}$ in complete disagreement with the experimental observations.

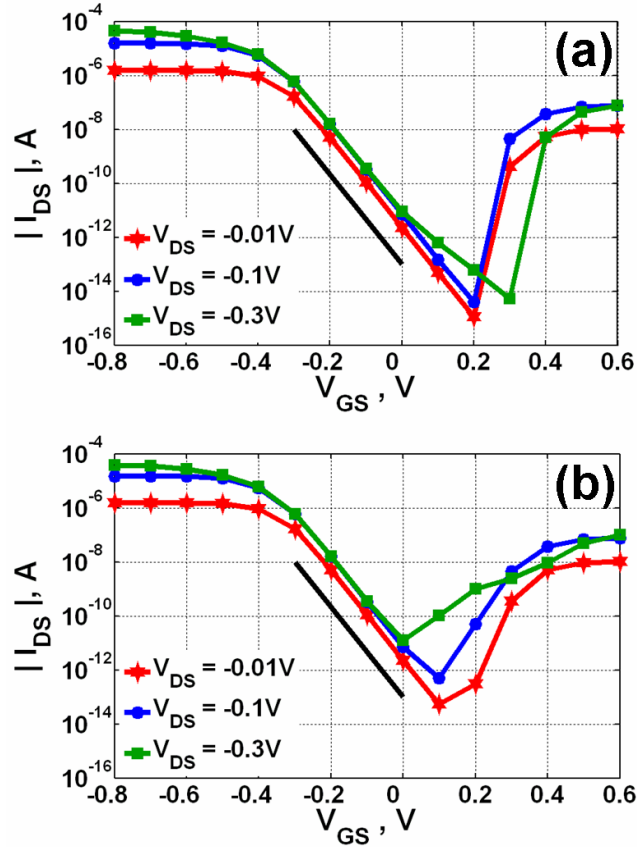

Figure 4. Simulated $I_{DS}$-$V_{GS}$ characteristics for the p-type CNT-MOSFET (black solid line - 60mV/dec swing): (a) ballistic, (b) one-phonon scattering with 195meV LO mode.

Figure 4(b) shows the simulated $I_{DS}$-$V_{GS}$ results for the model p-type CNT-MOSFET in the presence of phonon scattering. The sub-threshold swing at $V_{DS}$ = -0.01V for the steepest section of the BTBT transport is $S \approx 35$mV/dec, which is poorer compared to the ballistic case.[16] It is also observed to further degrade *with increasing $V_{DS}$*, in agreement with the experimental observations in Figure 2. Also, *the onset of BTBT transport occurs at more negative $V_{GS}$ (moves to the left) with increasing $V_{DS}$*, which is also in agreement with the experimental data. This shift in the onset of BTBT to lower voltages is due to phonon-assisted transport starting at more negative gate biases for larger $V_{DS}$. It is also important to note that in the presence of optical phonons, the charge pile-up effect discussed earlier is suppressed since the conduction band states are constantly emptied into the source due to inelastic tunneling (see Figure 3(b)). At this point, by comparing the experimental data with ballistic *I-V* results versus that with optical phonon scattering, it is clear that the former completely fails to explain the observed BTBT features, while the inelastic transport simulation seems to be essential for the correct description. The value for the on-current in BTBT operation (at $V_{GS}$ = 0.6V) with and without scattering, however, is very similar since at large positive gate biases the device current is mainly determine by the direct tunneling component. Thus, it is apparent that the BTBT operation of CNT-MOSFETs is governed by phonon-assisted transport, and dominates the sub-threshold characteristics in this regime. At this point it should be noted that elastic scattering due to acoustic phonons is not expected to affect the BTBT sub-threshold swing which is mainly determined by the inelastic tunneling due to optical phonons. The saturation current in the BTBT regime, however, can be reduced by acoustic phonon back-scattering when such events become allowed by energy conservation. It should also be pointed out that our simulations show slight reduction of the BTBT on-current at large gate biases, $V_{GS} > 0.6$V. This reduction is, however, smaller than that seen in Fig. 2 at low drain biases.

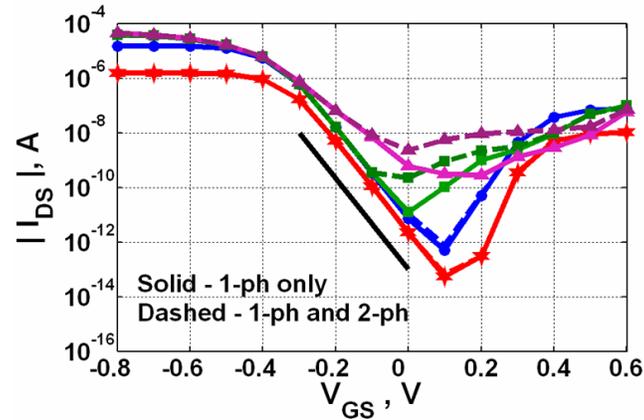

Figure 5. Simulated $I_{DS}$-$V_{GS}$ with only one-phonon scattering due to 195meV zone center LO mode (solid lines) and, also including two-phonon scattering due to 2 x 180meV zone-boundary mode (dashed lines). Red-star, blue-circle, and green-square curves are at the same voltages as in Figure 4(b), magenta-triangle curves are at $V_{DS} = -0.6$V.

The device operation under large drain biases is discussed next. The solid-triangle curve of Figure 5 shows the $I_{DS}$-$V_{GS}$ relationship for the model p-type CNT-MOSFET at $V_{DS} = -0.6$V under 195meV LO phonon scattering. It is seen that, due to the charge pile-up effect at such a high drain bias, the onset of BTBT is pushed to larger positive voltages (right). On the other hand, Figure 2 indicates that the experimentally observed onset of BTBT does not show such effects at large drain biases, and may move further left to smaller gate voltages. Examining Figure 3(b), it is clear that the onset of BTBT can be moved left to smaller gate biases if there are higher energy optical phonons that could allow inelastic tunneling. The highest energy optical phonons available in CNTs, however, are the ~195meV LO mode. On the other hand, Raman scattering experiments on CNTs report strong evidence for multi-phonon mediated processes (overtones).[31,32] For example, the $G'$-band, attributed to two-phonon Raman scattering involving the zone-boundary optical phonons, is observed to have similar intensities compared to the $G$-band arising from the zone-center LO phonons.[31,32] The $I_{DS}$-$V_{GS}$ results obtained with inclusion of two-phonon scattering are shown by the dashed curves in Figure 5. One can see that at small drain biases, it produces a negligible difference due to Pauli blocking of the two-phonon scattering mechanism. At larger drain biases (exceeding the energy of two phonons), this mechanism becomes effective. For such drain biases the device current near the onset of BTBT is dominated by two-phonon assisted tunneling due to one-phonon process being energetically inactive, and, the relatively strong e-ph coupling for multi-phonon mediated processes in CNTs, a fact observed in Raman experiments.[32] In Fig. 5 the onset of BTBT is indeed moved to lower voltages (i.e. moves left) compared to the one-phonon case, and the overall $I$-$V$ characteristics look even closer to the experimental data. Thus, it can be concluded that at large drain biases multi-phonon assisted inelastic tunneling might also become important for BTBT transport in CNT-MOSFETs.

In conclusion, we have performed detailed simulations for BTBT operation of CNT-MOSFETs using both ballistic as well as dissipative quantum transport. By comparing the simulation results with the experimental data, we conclude that the BTBT regime is dominated by optical phonon-assisted inelastic transport. It also appears that under large biasing conditions, multi-phonon scattering may also become important. The strong effect of optical-phonons on BTBT transport should be contrasted with conventional CNT-MOSFET operation, where their influence is found to be marginal up to moderate biases.[33,34] It is observed that BTBT operation can indeed produce sub-threshold swings below the conventional limit of 60mV/dec, which makes these devices attractive for low-power applications. The sub-threshold properties, however, are found to severely degrade under typical biasing conditions, and sensitively depend on phonon energies, device geometry, source/drain doping, etc.[16]

**Acknowledgment** - The authors acknowledge the support of this work by the NASA Institute for Nanoelectronics and Computing (NASA INAC NCC 2-1363) and Intel Corporation. Computational support was provided by the NSF Network for Computational Nanotechnology (NCN). S.O.K thanks fruitful communications with Xinran Wang of Stanford University and the Intel Foundation for PhD Fellowship support.